\documentclass[nofootinbib]{revtex4}
\usepackage{tikz-feynman}
\usepackage[english]{babel}
\usepackage{array,booktabs} 
\usepackage{array} 
\usepackage{relsize}
\usepackage{calc}
\usepackage{pdflscape}
\usepackage{color}
\usepackage{graphicx}
\usepackage{amsmath}
\usepackage[nodisplayskipstretch]{setspace}

\usepackage{setspace}
\usepackage{tabularx}

\usepackage{graphics}
\usepackage{amssymb}
\usepackage[font=small,labelfont=bf]{caption}
\usepackage{graphicx}
\usepackage{epstopdf}
\usepackage{appendix}
\usepackage{soul}
\usepackage{color}
\usepackage{subcaption}
\definecolor{blizzardblue}{rgb}{0.67, 0.9, 0.93}
\definecolor{bubblegum}{rgb}{0.99, 0.76, 0.8}
\usepackage[urlcolor=blizzardblue]{hyperref}
\hypersetup{
  colorlinks=false,
  linkcolor=blue,
  filecolor=magenta,    
  urlcolor=bubblegum,
}
\usepackage{makecell}
\usepackage{adjustbox}

\begin{document}
\title{\boldmath Scalar Dark Matter and Stability of Higgs Vacuum within a Minimal SO(10) GUT Model}
\author{ Nilavjyoti Hazarika\footnote{E-mail: hnilavjyoti@gmail.com}, Paramita Deka \footnote{deka.paramita@gmail.com}, Kalpana Bora\footnote{kalpana@gauhati.ac.in}} 

\affiliation{Department Of Physics, Gauhati University, Assam, India}

\begin{abstract}
In this work, we delve to investigate a feasible range of dark matter (DM) masses within a non-supersymmetric $SO(10)$ Grand Unified Theory (GUT) scalar dark matter model, in freeze-out scenario. This model includes a singlet scalar denoted as S and an inert doublet represented by $\phi$. Being part of SO(10), the quantum numbers of DM particles are assigned and hence we know their nature. These fields are odd under a discrete $Z_2$ matter parity $(-1)^{3(B-L)}$. The dark matter with mass 300 $\leq M_{DM} \leq$ 1000 GeV emerges as a mixture of the $Z_2$-odd scalar singlet $S$ and the neutral element of the doublet $\phi$, both residing within a \textbf{16}-dim scalar representation of $SO(10)$. Though similar propositions are available in literature for complex singlet scalar \textbf{S} as DM, in this work we consider a real scalar \textbf{S} belonging to \textbf{16} of SO(10) as DM.  We also investigate the one-loop vacuum stability through the solution of Renormalization Group Equations (RGEs) for the model's parameters. Subsequently, we scrutinize the model's predictions within the confines of contemporary theoretical and experimental restrictions. The novelty of this work is that, in addition to achieving vacuum stability in the SO(10) framework, the DM mass is found to lie in the previously unaddressed (theoretically) intermediate mass region of 300$\leqslant$ MDM $\leqslant$ 1000 GeV which also adheres to many current phenomenological constraints, like recent direct experimental bounds from XENON1T, indirect detection bounds from Fermi-LAT experiment, Higgs invisible decay and Electroweak Precision Test. The stability of the electroweak vacuum is seen to be present up to the Planck scale. These model predictions possess the potential for future validation through dark matter search experiments, as they are testable in future DM search experiments, along with the added feature that the model is a part of an elegant grand unified theory. 
\end{abstract}
\maketitle

\section{Introduction}\label{sec:level1}
Following the discovery of the Higgs boson by ATLAS and CMS experiments at the LHC in 2012, the value of the Higgs mass is determined accurately to be around 125 GeV \cite{ATLAS:2012yve, CMS:2012qbp}. Subsequently, measurements \cite{ATLAS:2013xga, CMS:2014nkk, CMS:2018uag, ATLAS:2019nkf} on its properties such as spin, parity, and interactions with SM particles validated the SM at energies accessible to the LHC. Although the standard model (SM) of particle physics is one of the most successful theories; however, it still has a few unresolved observations. The shortcomings of the SM are mainly- the strong CP problem, neutrino mass and mixing, matter-antimatter asymmetry, nature of dark matter (DM) and dark energy. Astrophysical and cosmological observations indicate that about 26 $\%$ of the current universe's energy density is composed of DM \cite{ParticleDataGroup:2018ovx}, with a present abundance expressed as $\Omega_{DM}h^{2}=0.1198\pm0.0026$ \cite{Planck:2018vyg}, where $\Omega_{DM}$ represents the DM density parameter and $h=\boldsymbol{H}/100$ $km s^{-1}Mpc^{-1}$, where $\boldsymbol{H}$ denotes the Hubble parameter. 

The SM also fails to provide absolute stability for the electroweak (EW) vacuum up to the Planck scale \cite{Isidori:2001bm, Bezrukov:2012sa, Buttazzo:2013uya}, with the EW vacuum residing in a metastable state instead of a stable state within the SM framework. Having known the values of Higgs mass, the top quark mass to be about 176 GeV \cite{CDF:1995wbb} and the value of the Higgs vacuum expectation value (VEV) to be $v = 246$ GeV, the SM vacuum at energy scale around $10^{10}$ GeV starts to be metastable \cite{Elias-Miro:2011sqh, Tang:2013bz, Degrassi:2014hoa, Rojas:2015yzm}. This is due to the fact that in the SM, the top quark has a large negative contribution in the renormalisation group equations (RGE) for the Higgs quartic coupling $\lambda_h$. Thus, the Higgs quartic coupling becomes negative at higher energy scales which in turn makes the vacuum metastable. Given these limitations, it is necessary to introduce new physics beyond the SM (BSM) as the observed instability implies limitations in its description of particle interactions and the behaviour of the Higgs field.

It is known that in the scalar singlet DM model, the only region which is not yet excluded is a narrow region close to the Higgs resonance $m_s\sim (m_h/2)$, others ruled out from different experimental and theoretical bounds (see \cite{Hazarika:2022tlc} and references therein). In the inert doublet model, the mass region ( 60-80 GeV) and the high-mass region (heavier than 550 GeV) are allowed. This motivated us to explore a parameter range in the so far theoretically unaddressed DM mass region, the intermediate-mass region $M_W <M_{DM} < 550$ GeV, which we successfully did in a scotogenic extension of SM with a scalar doublet and scalar singlets  \cite{Hazarika:2022tlc}.

One possibility to address the issue of vacuum stability is the addition of scalars, so that the vacuum can become stable up to the Planck scale. It has been extensively studied in the literature. For instance, extending the SM with an extra real scalar with $Z_2$ symmetry is studied
in \cite{Gonderinger:2009jp, Falkowski:2015iwa}, with a complex scalar in \cite{Elias-Miro:2012eoi, Gonderinger:2012rd, Athron:2018ipf}, Two-Higgs doublet models (2HDM) in \cite{Ferreira:2004yd, Maniatis:2006fs, Battye:2011jj, Kannike:2016fmd, Xu:2017vpq}. Further, the vacuum stability is studied in type-II seesaw models with $SU(2)_L$ -triplet scalars in \cite{Gogoladze:2008gf, Chun:2012jw, BhupalDev:2013xol, Bonilla:2015eha, Haba:2016zbu, Dev:2017ouk}, in $U(1)$ extensions of SM \cite{Datta:2013mta, Coriano:2014mpa, Haba:2015rha, Das:2015nwk}, as well as in the left-right symmetric models \cite{Mohapatra:1986pj, BhupalDev:2018xya, Chauhan:2019fji}.

Though one can propose many extra BSM fields in the theory, however, the question remains - from where these fields originate in a single theory? And, the answer to this question lies with Grand Unified Theories (GUTs), that  are very lucrative proposals, as they can accommodate solutions to many problems, in a holistic manner. Therefore, in view of the above mentioned issues, in this work we have undertaken to address the EW vacuum stability problem along with the existence of DM in intermediate mass region (300$\leq M_{DM}\leq$1000 GeV), as a manifestation of a GUT. In the GUT framework, in addition to explaining the origin of DM, it is also possible to  determine the nature of the DM particle and constrain its properties, due to its pinpointed quantum numbers. We know that in a GUT theory, at very high energy, all three interactions of the SM exhibit identical strength. However, as a result of the distinct renormalization group evolution with energy driven by the particle content of the model, they ultimately manifest varying strengths at lower energy levels. Any prospective GUT candidate must include the SM gauge group as a subgroup, thereby ensuring that the SM remains applicable at the electroweak scale. Since the SM's gauge group has a rank of four, the GUT's gauge group must possess a rank greater than or equal to four. These attributes are naturally realized in SO(10) GUTs \cite{Georgi:1974my, Fritzsch:1974nn, Chanowitz:1977ye, Georgi:1979ga, Mohapatra:1982aq, Babu:1992ia}, that offer numerous appealing characteristics, including:

\begin{itemize}
\item First, it allows for the incorporation of all SM quarks, leptons, and right-handed neutrinos within \textbf{16} representations of SO(10).
\item Second, it naturally resolves the issue of anomaly cancellation in the SM, as it is devoid of anomalies.
\item Third, achieves partial unification at an intermediate mass scale within SO(10) which results in improved gauge coupling unification \cite{Georgi:1979ga, Masiero:1980dd} and improved fermion mass ratios \cite{Georgi:1979ga, Lazarides:1980nt}.
\item Fourth - It can accommodate DM particle.
\end{itemize} 

In SO(10) GUT models, stable DM could be incorporated in a straightforward way \cite{Frigerio:2009wf, Kadastik:2009cu, Kadastik:2009dj, Mambrini:2013iaa, Mambrini:2015vna, Nagata:2015dma}. Additionally, SO(10) incorporates an extra $U(1)$ symmetry, which is assumed to undergo breaking at the intermediate scale. Multiple breaking chains for breaking SO(10) down into the SM are possible, aligning with experimental observations while also offering unique predictions. SO(10) introduces new particles beyond the SM, and may modify the running of the SM Higgs quartic coupling $\lambda_H$ through the addition of new particles at an intermediate scale below $10^{10}$ GeV, ensuring that $\lambda_H$ remains positive all the way up to the Planck scale \cite{Mambrini:2016dca}. In earlier works it has been shown that the existence of DM is a possible consequence of SO(10) GUT symmetry breaking. Gauge coupling unification and dark matter production through non-equilibrium thermal processes in the context of SO(10) models was presented in \cite{Mambrini:2013iaa}. In \cite{Frigerio:2009wf}, TeV scale fermion isotriplet with zero hypercharge belonging to a \textbf{45} (or larger) representation of SO(10) was considered to be DM candidate. In \cite{Kadastik:2009cu, Kadastik:2009dj}, a minimal non-supersymmetric SO(10) framework was considered with DM belonging to complex scalar singlet and doublets both belonging to \textbf{16}, where it was shown that electroweak symmetry breaking occurs radiatively due to DM couplings to the SM Higgs boson. In \cite{Mambrini:2015vna}, multiple breaking schemes of SO(10)  which lead to gauge coupling unification and stable DM along with production of light neutrino mass were presented. Scalar and fermion DM candidates in SO(10) GUT model is discussed in \cite{Nagata:2015dma}. In \cite{Mambrini:2016dca}, vacuum stability in addition to radiative electroweak symmetry breaking was studied in the SO(10) model, where a dark matter mass region of ($1.35-2$) TeV was found to be consistent with the theoretical and experimental bounds. In \cite{Sahu:2022rwq}, they considered an extension of SM by a scalar leptoquark and a fermion triplet embedded in a non-SUSY SO(10) GUT, and  the neutral component of fermion triplet served as DM candidate and unification mass scale along with the corresponding proton decay constraints were discussed.

The scalar DM models being simple and attractive, in this work, we address two of the important problems discussed above, that remain unaddressed so far - the scalar DM is the intermediate mass region and vacuum stability, in a single theory. We choose a minimal DM model where the DM is an admixture of a real SM singlet scalar that arises from a \textbf{16} representation and an Inert Doublet $\phi$ that belong to same \textbf{16} representation of the SO(10) GUT model. The novelty of this work is that, in addition to achieving vacuum stability in the SO(10) framework, the DM mass is found to lie in the intermediate mass region of 300 $\leq M_{DM} \leq$ 1000 GeV which also adheres to many current phenomenological constraints, like recent direct experimental bounds from XENON1T \cite{XENON:2017vdw, XENON:2018voc}, indirect detection bounds from Fermi-LAT experiment \cite{Fermi-LAT:2015att}, Higgs invisible decay \cite{ATLAS:2015ciy} and Electroweak Precision Test as well \cite{Baek:2012uj}. All these issues have not been addressed previously in a single model, and that makes the work interesting.

\par The paper is structured as follows: In section \ref{sec:1}, we describe our model. In section \ref{sec:2}, we present the theory for computing DM abundance and the DM scattering cross-section in the  the freeze-out scenario. In section \ref{sec:3}, we discuss the method used to achieve electroweak vacuum stability. For completeness, we give the expressions for one-loop beta functions used in our analysis in this section. Section \ref{sec:4} contains our results, which are subsequently discussed. Finally, in section \ref{sec:5}, we present a summary of our work and our conclusions. 

\section{The Model}
\label{sec:1}
As stated earlier,  in this work we aim to solve the problem of  stability of vacuum as well to find the DM candidate in so far theoretically unaddressed intermediate mass range, in SO(10)  grand unified theory. Our objective is to elucidate the nature and origin of DM, as well as constrain the parameter space based on current experimental limitations. Additionally, we wish to examine the stability of the electroweak vacuum up to the Planck scale, all within the SO(10) GUT framework. To stabilize the DM, we must enforce a discrete $Z_2$ symmetry. However, Planck scale operators break global discrete symmetries, which can be remedied by obtaining a $Z_2$ symmetry through the breaking of a gauged $U(1)$ embedded within a GUT. This could be achieved by breaking SO(10) GUT group to the SM gauge group and an extra $ U(1)_{X}$ subgroup, where $X$ is orthogonal to the SM hypercharge $Y$. Therefore, SO(10) can be broken down to the symmetry group of the SM and the gauged $Z_{2}$ parity \cite{Kadastik:2009cu}
\begin{equation}
P_{X}\equiv P_{M}=(-1)^{3(B-L)}
\end{equation}
which is the matter parity. We assume that the SO(10) gauge group is spontaneously broken to an intermediate subgroup $G_{int}$ at the GUT scale $M_G$, and subsequently broken to the SM gauge group $G_{SM} \equiv SU(3)_C \times SU(2)_L \times U(1)_Y $ and $P_{M}$ at an intermediate EWSB scale $M_Z$ :
\begin{equation}
SO(10) \rightarrow G_{int} \rightarrow G_{SM}\times P_{M}
\end{equation}
Some possible intermediate subgroups $G_{int}$ are listed in the following table:

\begin{table}[h]
\begin{center}
\begin{tabular}{|c|c|c|}
\hline
$G_{int}$ \\
\hline
$SU(4)_C\times SU(2)_L\times SU(2)_R$\\
\hline
$SU(4)_C\times SU(2)_L\times SU(2)_R\times D$\\
\hline
$SU(4)_C\times SU(2)_L\times U(1)_R$\\
\hline
$SU(3)_C \times SU(2)_L\times SU(2)_R\times U(1)_{B-L}$\\
\hline
$SU(3)_C \times SU(2)_L\times SU(2)_R\times U(1)_{B-L} \times D$\\
\hline
$SU(3)_C \times SU(2)_L\times U(1)_R\times U(1)_{B-L}$\\
\hline
$SU(5)\times U(1)$\\
\hline
\end{tabular}
\end{center}
\caption{Candidates for $G_{int}$ (the intermediate gauge group).\cite{Mambrini:2015vna}}
\label{tab:31}
\end{table}

In the above table, $D$ denotes the so-called $D$-parity or the left-right symmetry \cite{Frigerio:2009wf}, that is, the symmetry with respect to the exchange of $SU(2)_L \leftrightarrow SU(2)_R$. $D$-parity can be related to an element of SO(10)  \cite{Kuzmin:1980yp, Kibble:1982dd, Chang:1983fu, Chang:1984uy} under which a fermion field transforms into its charge conjugate.

Within the SO(10) GUT framework, the DM is postulated to be a combination of a $P_M$-odd scalar singlet $S$, which belongs to the $SU(2)_L \times U(1)_Y$ sector, and the neutral component of the doublet $\phi$, which is part of a new scalar \textbf{16} of SO(10). The grand unified group SO(10) undergoes a breakdown resulting in $SU(3)_c \times SU(2)_L \times U(1)_Y \times P_M$, where only the SM Higgs boson $H$ belongs to \textbf{10}, and the DM candidates - a mixture of the singlet S belonging to \textbf{16} and the Inert Doublet $\phi$ belonging to \textbf{16} - are light, while all other particles have masses of the order of GUT scale $M_G$. The SO(10) GUT breaking may occur in one or in several steps through intermediate subgroup as mentioned in table \ref{tab:31}. In this work we assume these breakings to occur near the GUT scale.

To verify the proposed DM scenario, we analyze the scalar potential of a minimal SO(10) GUT model consisting of a scalar \textbf{16} (i.e., $\phi$) for the DM and a scalar \textbf{10} (i.e., H) for the SM Higgs doublet. The inert doublet and SM Higgs are as follows:

\begin{equation}
H=\begin{pmatrix} 
0\\ \frac{1}{ \sqrt{2}}(v+h)
\end{pmatrix}
\quad
,
\phi=\begin{pmatrix} 
H^{+}\\ \frac{1}{ \sqrt{2}}(H_{0}+iA_{0})
\end{pmatrix}
\end{equation}\\

The $SO(10)$ symmetric scalar potential of one \textbf{16} and one \textbf{10} is \cite{Kadastik:2009cu}
\begin{eqnarray}
\begin{array}{rcl}
V &=& \mu_{1}^2 (\textbf{10}\hspace{0.1cm}\textbf{10}) + \mu_{2}^2 (\textbf{16}\hspace{0.1cm}\textbf{16})+\lambda_h (\textbf{10}\hspace{0.1cm}\textbf{10})^2 + \lambda_2 (\textbf{16}\hspace{0.1cm} \textbf{16})^2 + \lambda_3 ( (\textbf{10\hspace{0.1cm}10})(\textbf{16}\hspace{0.1cm}\textbf{16})\\&&
+ \lambda_4 (\textbf{16\hspace{0.1cm}10})(\textbf{16\hspace{0.1cm}10}))
 + \rho (\textbf{16\hspace{0.1cm}10\hspace{0.1cm}16} + hc)
\end{array}
\end{eqnarray}

For simplicity, we have assumed that all parameters are real. The scalar potential below $M_G$ that we consider is the most general CP-invariant potential that remains invariant under the transformations $H \rightarrow H$, $\phi \rightarrow \phi$, and $S \rightarrow -S$.

\begin{eqnarray}
\left. \begin{array}{rcl}\label{1}
V^{'}&=&\mu_{1}^{2}(H^{\dagger}H)+\mu_{2}^{2}(\phi^{\dagger}\phi)+\mu_{S}^{2}(S^{2})+\lambda_{h}(H^{\dagger}H)^{2}+\lambda_{2}(\phi^{\dagger}\phi)^{2}\\&&+\lambda_{S}(S)^{4}+\lambda_{3}(H^{\dagger}H)(\phi^{\dagger}\phi)+\lambda_{4}(H^{\dagger}\phi)(\phi^{\dagger}H)\\&&+\frac{\lambda_{5}}{2}[(H^{\dagger}\phi)^{2}+ h.c]+\lambda_{HS}(S^{2})(H^{\dagger}H)+\lambda_{\phi S}(S^{2})(\phi^{\dagger}\phi)\\&&+\rho[(H^{\dagger}\phi)S+h.c]
\end{array}\right.
\end{eqnarray}

where $H$ is the SM Higgs doublet (belonging to \textbf{10}), and \textbf{16} contain the inert doublet $\phi$ and the singlet $S$. If $M_G$ is the GUT unification scale, the GUT scale boundary conditions can be expressed as, \cite{Kadastik:2009cu}

\begin{equation}\label{gutbc}
\begin{aligned}
\mu^2_{1}(M_G)>0, \quad \mu^2_{2}(M_G)=\mu^2_{S}(M_G) >0;\\
\lambda_{2}(M_G) = \lambda_{S}(M_G)= \lambda_{\phi S},\quad \lambda_{3}(M_G) =\lambda_{H S}(M_G)\\
\end{aligned}
\end{equation}

We consider mixing between the real singlet $S$ and Inert doublet $\phi$ and obtain a $2\times2$ mass matrix for the lightest neutral $Z_{2}$ odd scalar as \cite{Hazarika:2022tlc}
\begin{eqnarray}
M=\begin{pmatrix}
\mu_{2}^{2}+\lambda_{L}v^{2} & \rho v \\
  \rho v & 2\mu_{S}^{2}+\lambda_{HS}v^{2}
 \end{pmatrix}=\begin{pmatrix}
m_{H_{0}}^{2} & m_{H_{0}S}^{2}\\
  m_{H_{0}S}^{2} & m_{S}^{2}
 \end{pmatrix}
\end{eqnarray}
\vspace{0.2cm}

where {\large $\lambda_{L}=\frac{\lambda_{3}+\lambda_{4}+\lambda_{5}}{2}$}. The physical mass eigenstates can be defined as
\begin{eqnarray}
\begin{array}{l}
\chi_{1}=H_{0}\cos \theta + S \sin \theta \\
\chi_{2}=-H_{0}\sin \theta + S \cos \theta
\end{array}
\end{eqnarray}\\
Masses of the neutral physical scalars $\chi_{1}$ and $\chi_{2}$ are,\\
\begin{eqnarray}\label{2}
\begin{array}{l}
m_{1}^{2}=\frac{m_{H_{0}}^{2} + m_{s}^{2}}{2} - \frac{m_{H_{0}}^{2}-m_{s}^{2}}{2}\sqrt{1+\tan^{2}2\theta},\\
m_{2}^{2}=\frac{m_{H_{0}}^{2}+ m_{s}^{2}}{2} +\frac{m_{H_{0}}^{2}-m_{s}^{2}}{2}\sqrt{1+\tan^{2}2\theta}
\end{array}
\end{eqnarray}
\vspace{0.2cm}

The couplings in terms of the masses are ( see \cite{Hazarika:2022tlc})
\begin{eqnarray}
\begin{array}{l}
\lambda_{1}=\dfrac{\mu_{1}^{2}}{2v^{2}},\\

\lambda_{3}=\frac{2(\lambda_{L}v^{2}+\mu_{H^{\pm}}^{2}-m_{1}^{2}\cos^{2}\theta-m_{2}^{2}\sin^{2}\theta)}{v^{2}},\\

\lambda_{4}=\frac{m_{1}^{2}\cos^{2}{\theta}+m_{2}^{2}\sin^{2}{\theta}+\mu_{A_{0}}^{2}-2\mu_{H^{\pm}}^{2}}{v^{2}},\\

\lambda_{5}=\frac{m_{1}^{2}\cos^{2}{\theta}+m_{2}^{2}\sin^{2}{\theta}-\mu_{A_{0}}^{2}}{v^{2}},\\

\mu_{S}^{2}=\frac{1}{2}(m_{1}^{2}\sin^{2}{\theta}+m_{2}^{2}\cos^{2}{\theta}-\lambda_{HS}v^{2}),\\

\rho =\frac{(m_{2}^{2}-m_{1}^{2})\sin{2\theta}}{2v},\\

\mu_{2}^{2}=m_{1}^{2}\cos^{2}{\theta}+m_{2}^{2}\sin^{2}{\theta}-\lambda_{L}v^{2}

\end{array}
\end{eqnarray}

To ensure that the potential in our model remains bounded from below, vacuum stability necessitates \cite{Kadastik:2009cu}
\begin{equation}
 \begin{aligned}
\lambda_{h}, \lambda_{\phi}, \lambda_{S} >0;\quad \lambda_{3}+2\sqrt{\lambda_{h}\lambda_{2}} >0\\
\lambda_{3} + \lambda_{4} - |\lambda_{5} | + 2\sqrt{\lambda_{h}\lambda_{2}} >0\\
\end{aligned}
\end{equation}

\section{Dark Matter}\label{sec:2}
Now, we present a discussion on various constraint checks on dark matter as done in our analysis. 
\subsection{Relic density constraints}
As previously mentioned, the chosen DM candidate in our model is the lightest component, denoted as $\chi_{1}$, which is a mixture of singlet and doublet scalar DM fields. To ascertain the thermal relic abundance of this DM particle, which follows the WIMP archetype, we solve the Boltzmann equation \cite{Kolb:1990vq}

\begin{equation}
\frac{dn_{DM}}{dt}+3 \boldsymbol{H} n_{DM}=-<\sigma v>[n_{DM}^{2}-(n_{DM}^{eq})^{2}]
\end{equation}

where $n_{DM}$ is the number density of the DM particle and $n_{DM}^{eq}$ is the number density of the DM particle when it is in thermal equilibrium. $\boldsymbol{H}$ is the Hubble rate of expansion of the Universe and $<\sigma v>$ is the thermally averaged annihilation cross-section of the DM particle. One can obtain the numerical solution of the Boltzmann equation in terms of partial expansion $ <\sigma v> = a +bv^{2}$ as \cite{Scherrer:1985zt}

\begin{equation}
\Omega_{DM}h^{2}\approx\frac{1.04\times 10^9 x_{F}}{M_{Pl}\sqrt{g_{\ast}}(a+\frac{3b}{x_{F}})}
\end{equation}

where $ x_{F}=m_{DM}/T_{F}$, $T_{F}$ is the freeze-out temperature, $g_{\ast}$ is the number of relativistic degrees of freedom at the time of freeze-out. After further simplifications, the above solution takes the form as \cite{Jungman:1995df}

\begin{equation}
\Omega_{DM}h^{2}\approx\frac{3\times 10^{-27} cm^{3}s^{-1}}{<\sigma v>}
\end{equation}

\subsection{DM-nucleon cross section (DD)}
The thermal averaged annihilation cross section $<\sigma v>$ is given by \cite{Gondolo:1990dk}

\begin{equation}
<\sigma v> =\frac{1}{m_{DM}^{4} T K_{2}^{2}(\frac{m_{DM}}{T})}\int \sigma (s-4 m_{DM}^{2})\surd s K_{1}(\frac{\surd s }{T})ds
\end{equation}

Here, $K_{i}$ represents the modified Bessel functions of order $i$, $m_{DM}$ is the DM particle's mass, and $T$ corresponds to the Universe's temperature. Moreover, the relevant spin-independent scattering cross-section concerning the scalar DM $H_{2}$, mediated through the SM Higgs, can be formulated as follows \cite{Barbieri:2006dq}:

\begin{equation}
\sigma_{SI}=\frac{\lambda_{L}^{2}f^{2}}{4\pi}\frac{\mu^{2}m_{n}^{2}}{m_{h}^{4}m_{DM}^{2}}
\end{equation}

Here, $m_{n}$ stands for the mass of a nucleon, and {\large $\mu=\frac{m_{n}m_{DM}}{m_{n}+m_{DM}}$} represents the reduced mass of DM and nucleons. Similarly, {\large $\lambda_{L}=\frac{\lambda_{3}+\lambda_{4}+\lambda_{5}}{2}$} signifies the quartic coupling between DM and the Higgs particle, while $f$ denotes the coupling between the Higgs and nucleons. According to a recent evaluation, $f$ has been approximated as $0.32$ \cite{Giedt:2009mr}, although the permissible range of values spans from $0.26$ to $0.63$ \cite{Mambrini:2011ik}.

\subsection{ DM-Higgs invisible decay}
\par Moreover, the DM-Higgs coupling $\lambda_{L}$ can be restricted based on the most recent constraints from the LHC regarding the hidden decay width of the SM Higgs boson. The invisible Higgs decay width has a connection to the DM-Higgs coupling $\lambda_{L}$ as explained in \cite{ATLAS:2015ciy} as,

\begin{equation}
\Gamma(h\rightarrow Invisible)=\frac{\lambda_{L}v^{2}}{64\pi m_h}\sqrt{1-4m_{DM}^{2}\diagup m_h^{2} }
\end{equation}

\subsection{Indirect detection (ID) and Fermi-LAT}
Apart from direct detection experiments, the realm of DM parameters can also be confined using findings from various indirect detection experiments, such as those carried out by Fermi-LAT \cite{Fermi-LAT:2015att}. These experiments focus on searching for SM particles, which are produced either through the annihilation of DM or through its decay in the local Universe. The resulting neutral and stable final products, including photons and neutrinos, can reach the indirect detection experiments with minimal interference from intermediate regions. For DM of the WIMP type, these photons fall within the gamma-ray range, detectable by space-based telescopes like Fermi-LAT through observations of dwarf spheroidal satellite galaxies (dSphs). The observed differential gamma-ray flux originating from DM annihilations is mathematically expressed as follows: \cite{Borah:2017dfn}

\begin{equation}
\dfrac{d\Phi}{dE}(\vartriangle\Omega)=\frac{1}{4\pi}<\sigma v>\frac{J(\vartriangle\Omega)}{2M_{DM}^{2}}\dfrac{dN}{dE}
\end{equation}
In this context, the portion of the sky under observation is represented by the solid angle $\vartriangle\Omega$, $<\sigma v>$ denotes the thermally averaged cross section for DM annihilation, and $dN/dE$ signifies the average gamma-ray spectrum produced per annihilation event. The astrophysical factor $J$ is mathematically expressed as follows: \cite{Borah:2017dfn}

\begin{equation}
J(\vartriangle\Omega)=\int_{\vartriangle\Omega}d\Omega'\int_{LOS}dl \rho^{2}(l,\Omega')
\end{equation}
where, $\rho$ represents the density of DM, and LOS stands for the line of sight. As a result, it becomes feasible to limit the DM annihilation into various final states, such as $\mu^{+} \mu^{-}, \tau^{+}\tau^{-}, W^{+} W^{-}, ZZ, b\bar{b}$. By utilizing the constraints derived from DM annihilation into these specific final states, we can confine the DM parameters through a comprehensive analysis of observations from Fermi-LAT \cite{Fermi-LAT:2015att} involving dwarf spheroidal satellite galaxies.

\section{Vacuum stability}\label{sec:3}
\begin{figure*}
\centering\includegraphics[width=10cm, height=7cm]{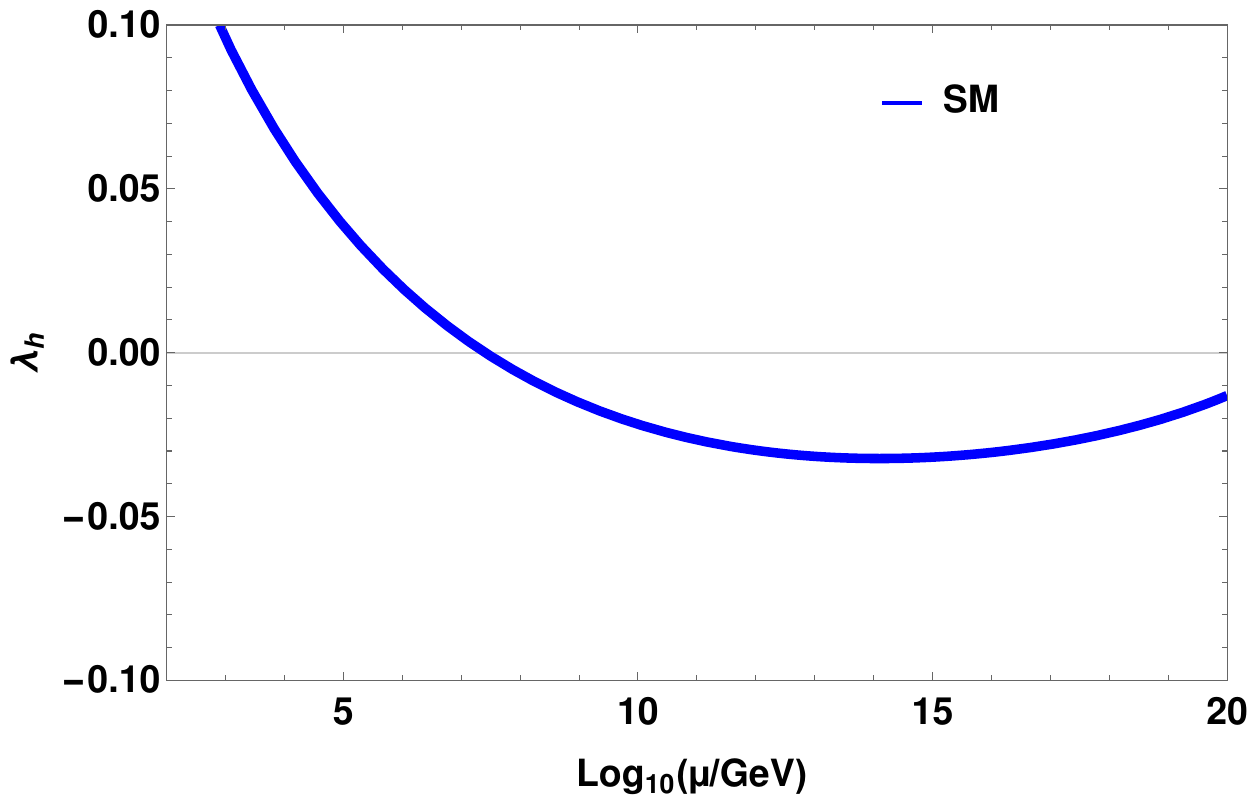}
\caption{Running of Higgs quartic coupling $\lambda_h$ in the SM}\label{fig:4.1}
\end{figure*}

The scalar potential associated with the SM Higgs doublet is
\begin{equation}
V_{SM} = \mu^{2} |H|^2 +\frac{\lambda_h}{2}|H|^4
\end{equation}
where, the parameters $\lambda_h$ and $\mu$ are influenced by both the Higgs Vacuum Expectation Value (VEV) and the Higgs particle's mass, denoted as $m_h$. If we assume the validity of the SM up to a very high energy scale, approximately $10^{10}$ GeV, an intriguing observation emerges. The quartic coupling $\lambda_h$ undergoes a transition to a negative value at energy scales below $10^{10}$ GeV as shown by the solid blue curve in figure \ref{fig:4.1}. This transition holds significant consequences. Specifically, the scalar potential within the SM becomes unbounded in the downward direction for sufficiently large values of the Higgs field ($H$). Consequently, the stability of the vacuum in the electroweak theory is compromised. These findings suggest the need for new physics to come into play at energy scales below $10^{10}$ GeV.

\begin{figure*}
\centering\includegraphics[width=10cm, height=7cm]{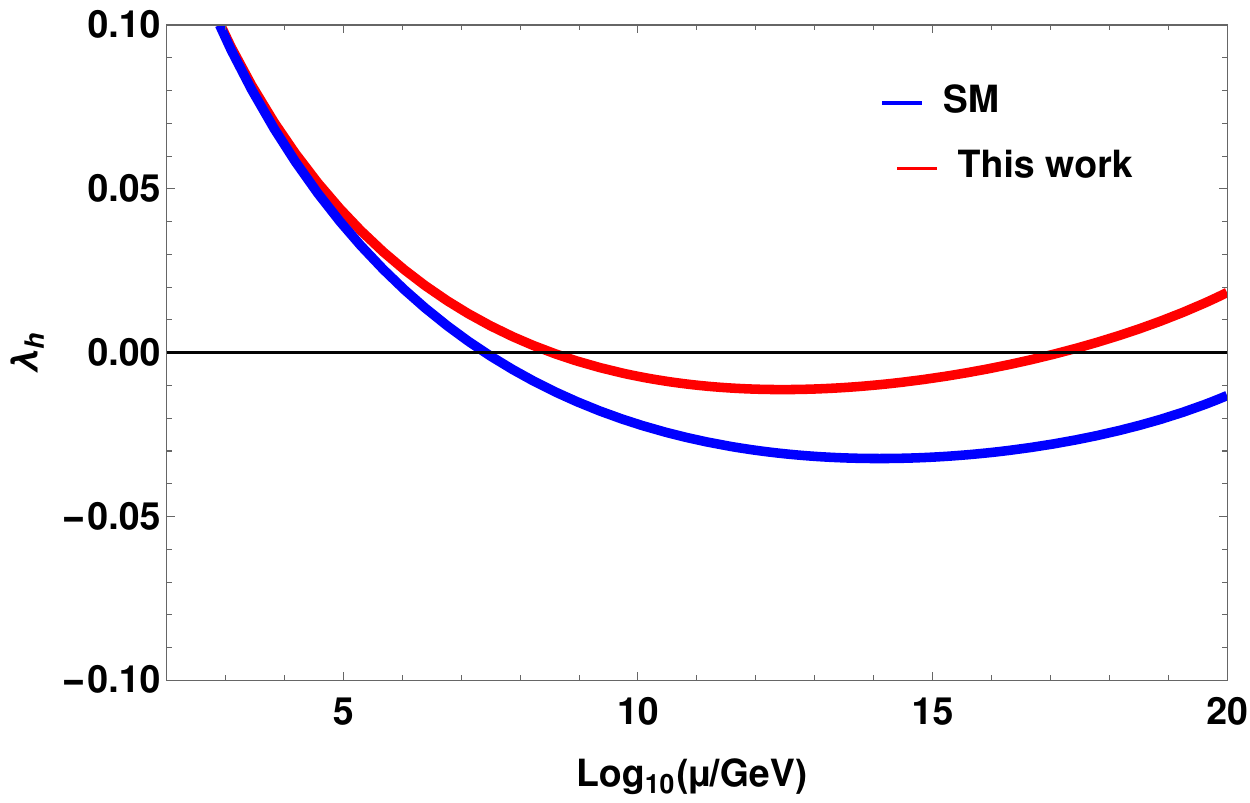}
\caption{Comparison of running of Higgs quartic coupling $\lambda_h$ in SO(10) with SM.}\label{fig:4.2}
\end{figure*}

Within the framework of an SO(10) unification model, fresh insights emerge. In this work we have considered the DM candidate to be a mixture of the singlet $S$ and neutral component of Inert Doublet $\phi$ belonging to its multiplet \textbf{16}. We use renormalization group equations (RGEs) for scalar mass parameters $\mu^{2}_{i}$ and interaction couplings $\lambda_{i}$ below the GUT breaking scale and study the vacuum stability for those parameters \cite{Ferreira:2009jb}. We also impose the GUT boundary conditions as stated in Eq.\ref{gutbc}. We use SARAH-4.14.3 package \cite{Staub:2013tta} to compute the RGEs. Here, a candidate for DM and an intermediate scale, lying below the energy scale of $10^{10}$ GeV, bring forth new particles that extend beyond the SM. The inclusion of new particles leads to an observation where the introduction of these particles causes the vacuum to exhibit a positive nature, extending all the way to the PLANCK scale $M_{Pl}$ as shown by the solid red curve in Figure \ref{fig:4.2}. 

The expressions of renormalization group equations (RGEs) for the model in Equation \ref{1} have been obtained, and the one-loop beta functions for the scalar couplings are as follows: \cite{Kadastik:2009cu}

\begin{equation} 
\beta_{\lambda_h} =  
24 \lambda_h^{2}+ 2\lambda_{3}^{2} +2 \lambda_{3}\lambda_{4}+\lambda_{4}^2+\lambda_{5}^2+\lambda_{SH}^2+\frac{3}{8}(3g_2^4+2g_2^2g_1^2)+g_1^4-3\lambda_{1}(3g_2^2+g_1^2-4y_t^2)-6y_t^4
\end{equation}

\begin{equation} 
\beta_{\lambda_2}=  
24\lambda_{2}^{2}+2 \lambda_{3}^{2}+2 \lambda_{3} \lambda_{4} + \lambda_4^2+ \lambda_{5}^2+ \lambda_{\phi S}^2 +\frac{3}{8}( 3g_{2}^4+2g_2^2 g_1^2) -3 \lambda_{2}( 3 g_2^2+ g_{1}^{2})
\end{equation}

\begin{equation} 
\beta_{\lambda_{3}}=  
4(\lambda_{h}+\lambda_{2})(3\lambda_{3}+\lambda_{4}) +4\lambda_{3}^{2}+2 \lambda_{4}^2 +2 \lambda_5^2 +2\lambda_{HS} \lambda_{\phi S}+\frac{3}{4}(3 g_2^4+g_1^4-2g_2^2 g_1^2)-3\lambda_{3}(3g_2^2+g_1^2-4y_t^2)
\end{equation}

\begin{equation}  
\beta_{\lambda_{4}}=  
4(\lambda_{h}+\lambda_{2})\lambda_{4}+ 8\lambda_{3} \lambda_{4} + 4\lambda_{4}^{2}+8 \lambda_{5}^2+3g_2^2 g_1^2-3\lambda_{4}(3g_2^2+g_1^2-4y_t^2)
\end{equation}

\begin{equation} 
\beta_{\lambda_{5}}=  
4(\lambda_{h}+\lambda_{2}+2\lambda_{3}+3\lambda_{4})\lambda_{5}-3\lambda_{5}(3g_2^2+g_1^2-2y_t^2)
\end{equation}

\begin{equation} 
\beta_{\lambda_{S}}=  
20 \lambda_{S}^2 +2 \lambda_{S}^2 +2\lambda_{\phi S}^2
\end{equation}
 
 \begin{equation} 
\beta_{\lambda_{HS}}=  
4(3\lambda_{h} +2\lambda_{S} +\lambda_{HS})\lambda_{HS} +(4 \lambda_3+2\lambda_{4})\lambda_{\phi S}-\frac{3}{2} \lambda_{SH}(3g_2^2+g_1^2-y_t^2)
\end{equation}

\begin{equation} 
\beta_{\lambda_{\phi S}}=  
4(3\lambda_{2} +2\lambda_{S} +\lambda_{\phi S})\lambda_{\phi S}+ (4 \lambda_3+2\lambda_{4})\lambda_{HS}-\frac{3}{2} \lambda_{\phi S}(3g_2^2+g_1^2)
\end{equation}

\begin{equation} 
\beta_{\mu_1^{2}} =  
12 \mu_1^2 \lambda_h +4\mu_2^2 \lambda_3 +2\mu_2^2 \lambda_{4} +2 \mu_S^2 \lambda_{HS} +\rho^2 -\frac{3}{2} \mu_{1}^2(3g_2^2+g_1^2-4y_t^2)
\end{equation}

\begin{equation} 
\beta_{\mu_2^{2}} =  
12 \mu_2^2 \lambda_2 +4\mu_1^2 \lambda_3 +2\mu_1^2 \lambda_{4} +2 \mu_S^2 \lambda_{\phi S} +\rho^2 -\frac{3}{2} \mu_{\phi S}^2(3g_2^2+g_1^2)
\end{equation}

\begin{equation} 
\beta_{\mu_S^{2}} =  
8 \mu_S^2 \lambda_S +4\mu_1^2 \lambda_{HS} +4\mu_2^2 \lambda_{\phi S} +2 \mu_S^2 \lambda_{\phi S} +\rho^2
\end{equation}

\begin{equation} 
\beta_{\rho}=  
2 \rho(\lambda_3 + 2 \lambda_{4} + \lambda_{HS}+\lambda_{\phi S}) -\frac{3}{2} \rho(3g_2^2+g_1^2-2y_t^2)
\end{equation}
Further we have included the one-loop $\beta$-functions for $g_1, g_2, g_3$ and $y_t$, given by
\begin{eqnarray}
\begin{array}{l}
\beta_{g_1}=7g^3_1\\
\beta_{g_2}=-3g^3_2\\
\beta_{g_3}=-7g^3_1\\
\beta{y_t}=y_t(\frac{9}{2}y^2_t-\frac{17}{12}g^2_1-\frac{9}{4}g^2_2-8g^2_3)
\end{array}
\end{eqnarray}
\section{Results and discussions}
\label{sec:4}

In this section, we explore how our SO(10) model implications relate to DM and contribute to the electroweak vacuum. We compare our results with the most recent limitations on DM relic density (derived from Planck data) and the spin-independent direct detection cross-section (measured by the XENON1T experiment). Additionally, we enforce a perturbativity condition ($\lambda_{i}$ $<$ 4$\pi$ across all scales) for all data points. Further, we enforce the Electroweak Precision Test (EWPT) restriction, as outlined in \cite{Baek:2012uj}, which establishes an upper limit on the mixing angles $\theta$ across the entire range of masses. These constraints are uniformly applied to all parameter points.

\begin{figure*}
\centering\includegraphics[width=10cm, height=7cm]{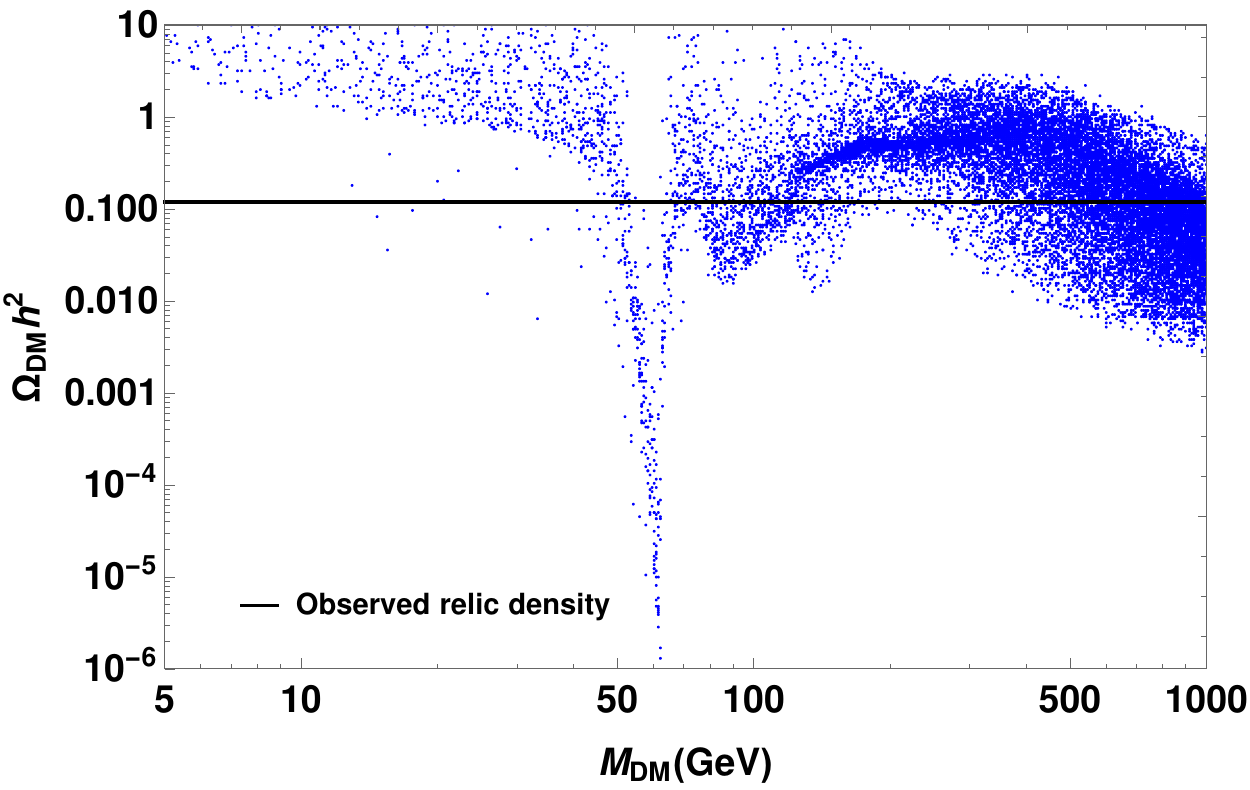}
\caption{The change in DM relic density $\Omega_{\text{DM}}h^{2}$ with respect to the mass of the DM is depicted for parameters $\lambda_{i}= 0.01$ and $\sin{\theta}=0.2$. The black line corresponds to the existing value of DM relic density. The region below the black line is the allowed region from relic density constraint.} \label{fig:2}
\end{figure*}

\begin{figure*}
\centering\includegraphics[width=10cm, height=7cm]{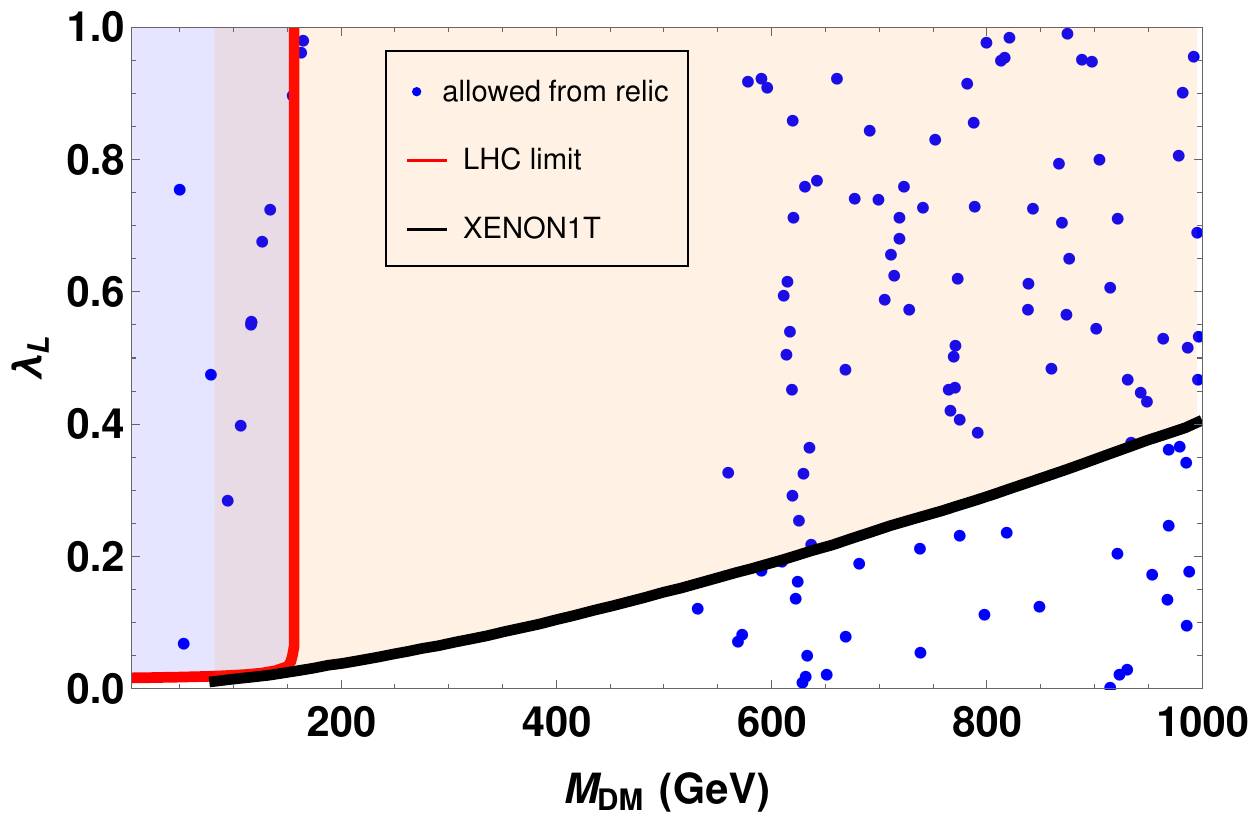}
\caption{Points that satisfy the relic density requirements are depicted in the $\lambda_{L}$ vs $M_{DM}$ plane. The blue points are allowed from correct relic density constraints. The exclusion lines in black and red correspond to the constraints from the XENON1T and LHC limits on Higgs invisible decay, respectively. The area to the right of the red line and below the black curve are allowed.}
\label{fig:3}
\end{figure*}

\begin{figure*} 
\centering\includegraphics[width=10cm, height=7cm]{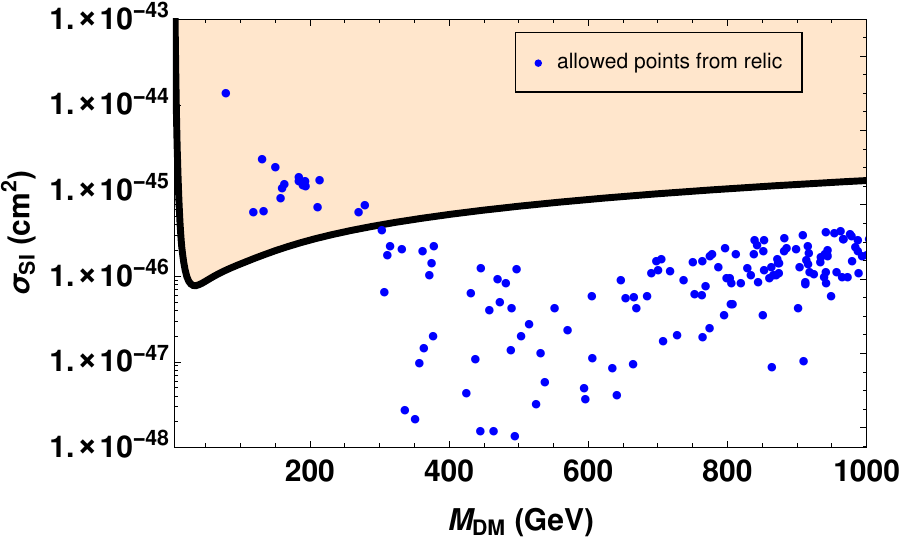}
\caption{The spin-independent cross-section of DM with nucleons, denoted as $\sigma_{SI}$ ($cm^{2}$), is presented for the points that satisfy the accurate relic density constraint. The exclusion curve resulting from the XENON1T experiment is represented by the black line. The region below the black curve are allowed from current constraints from XENON1T experiment.} \label{fig:4}
\end{figure*}

\begin{figure*}
\begin{subfigure}{0.48\textwidth}
\includegraphics[width=8cm, height=6cm]{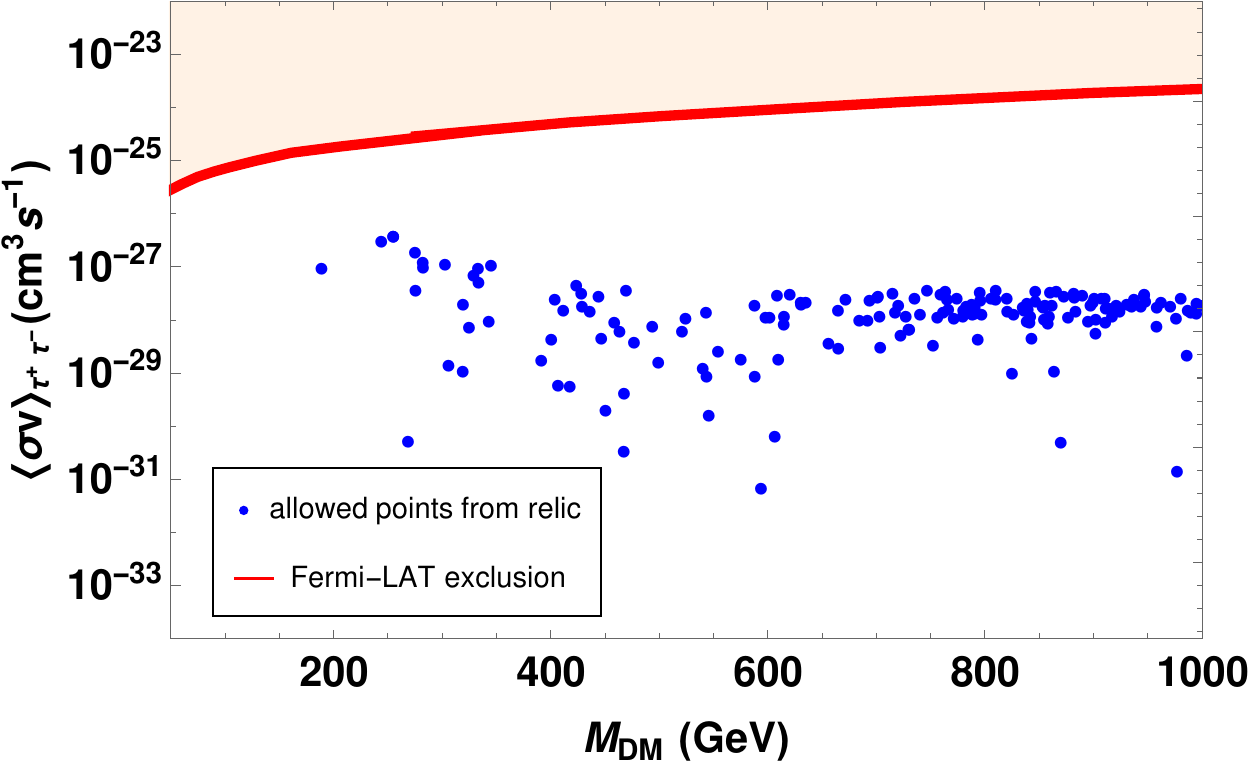}
\caption{DM annihilations into $\tau^{+}\tau^{-}$} \label{fig:5a}
\end{subfigure}\hspace*{\fill}
\begin{subfigure}{0.48\textwidth}
\includegraphics[width=8cm, height=6cm]{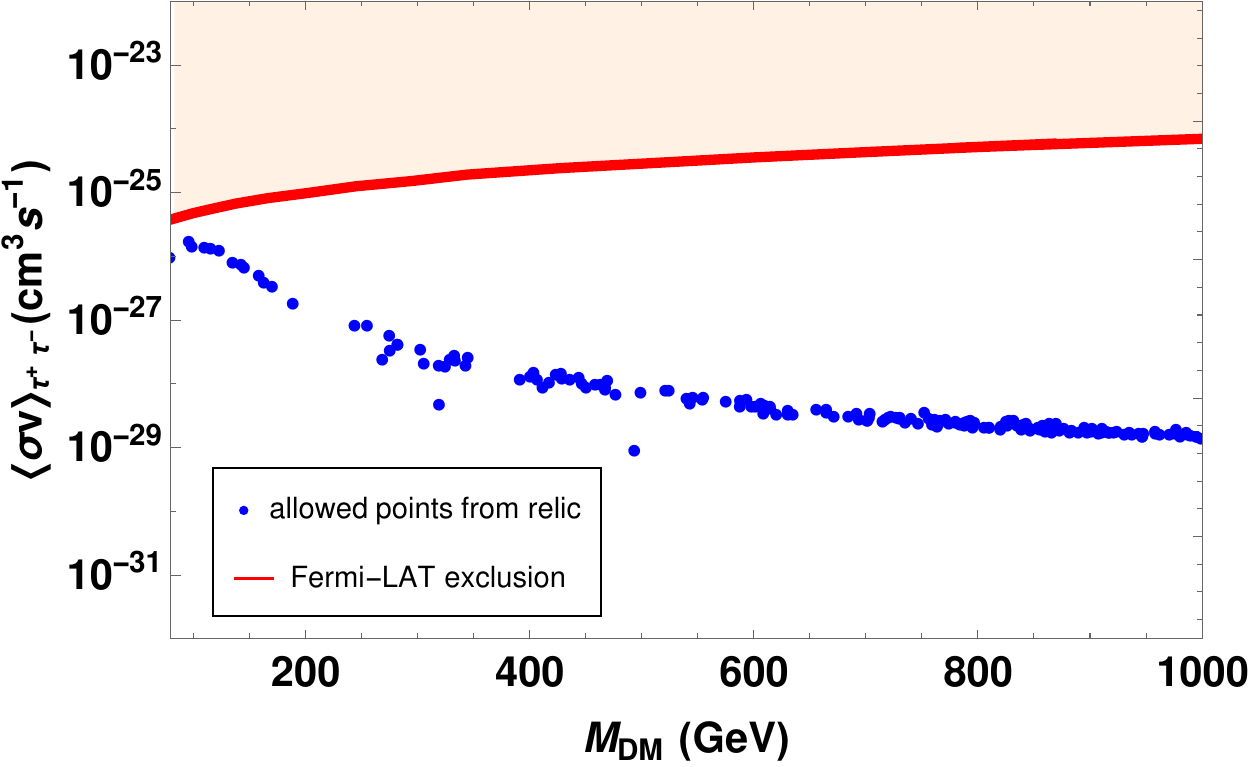}
\caption{DM annihilations into $W^{+}W^{-}$ } \label{fig:5b}
\end{subfigure}
\caption{DM annihilations into $\tau^{+}\tau^{-}$(left), $W^{+}W^{-}$ (right) compared against the latest
indirect detection bounds of Fermi-LAT. The regions below the red line are allowed.}\label{fig:5}
\end{figure*}

As stated earlier, we choose $\chi_{1}$ to serve as the dark matter candidate, which is a composite of singlet and doublet scalars. Subsequently, we explore the parameter space that adheres to the necessary constraints on relic density. We vary the DM mass in the range $5-1000$ GeV and use the software package micrOMEGA 4.3.2 \cite{Belanger:2013oya} to compute both the relic abundance and the spin-independent cross-section of the DM. The results of our study are presented in figures (\ref{fig:2}-\ref{fig:5}). In figure \ref{fig:2} we have shown the variation of the DM relic abundance with DM mass. It is seen that there is a significant area of parameter space that can produce the correct relic density of DM in our model. Additionally, a distinctive funnel-shaped area emerges around the Higgs resonance point ($M_{DM} \approx m_{h}/2$), corresponding to the s-channel annihilation of DM into SM fermions, facilitated by the Higgs boson mediation.

Subsequently, we check across viable values of the DM-Higgs coupling $\lambda_{L}$ for a given mass squared difference $(\bigtriangleup m= m_{\chi_{1}}^{2}-m_{\chi_{1}}^{2}= 50$ GeV), and show the permissible parameter range in the $\lambda_{L}$ vs $M_{DM}$ plane based on the requirement of achieving the accurate relic abundance, in Figure \ref{fig:3}. The exclusion lines in black and red correspond to the constraints from the XENON1T\cite{XENON:2017vdw} and LHC limits on Higgs invisible decay, respectively. The shaded region, lightly coloured in orange, is deemed impermissible due to the LHC limits \cite{ATLAS:2015ciy} with reference to invisible decay and the direct detection observations from the XENON1T experiment \cite{XENON:2018voc}. It is important to note that the latest LHC constraint on the invisible decay width of the SM Higgs boson is applicable solely for DM masses $m_{DM} < \frac{m_h}{2}$, which implies that the area to the left of the red line is not allowed. As a result, the parameter space region located to the right of the red line and situated below the black line successfully adheres to both LHC and XENON1T limitations. Consequently, it is reasonable to select smaller values of the coupling $\lambda_L \le 0.1$ in our analysis that follows.

Exploring the impact of varying coupling constants adds an intriguing dimension to our analysis. Thus, in figure \ref{fig:4}, we present the correlation between the spin-independent DM-nucleon cross-section $\sigma_{SI}$ ($cm^{2}$) and the coupling $\lambda_{L}$ values, which comply with the correct relic density constraint. We compare these results with the boundaries of the spin-independent DM-nucleon cross-section as constrained by the latest XENON1T experiment. As depicted in the spin-independent direct detection cross-section plot of figure \ref{fig:4}, it's evident that numerous parameter points within the space, satisfying the correct relic density constraint \cite{Planck:2018vyg}, also reside within the permissible range defined by the XENON1T experimental curve (situated below the black curve). Consequently, we deduce that the DM candidate with a mass of $M_{DM}\geq 300$ GeV and featuring a modest coupling $\lambda_L$ adheres to both the experimental constraints for the proper relic density and the DM-nucleon scattering cross-section.
Based on the outcomes illustrated in Figure \ref{fig:4}, it becomes evident that our model has the capability to predict DM candidates that meet both the constraints for relic density and cross-section. This validation of our model is particularly notable as it encompasses a wide range of DM masses, in previously unaddressed (theoretically) intermediate range.

Since the DM in our model consists of a combination of singlet and doublet scalars, it is important to recognize that aside from the constraints arising from direct detection experiments, the parameter space for DM can also be explored through various indirect detection experiments. We place restrictions on the DM parameters based on the indirect detection bounds derived from a comprehensive analysis of observations from Fermi-LAT of dwarf spheroidal satellite galaxies (dSphs) \cite{Fermi-LAT:2015att}. Figure \ref{fig:5} portrays the cross section for DM annihilation into final states $\tau^{+}\tau^{-}$ and $W^{+}W^{-}$, and the outcomes are compared with the most recent indirect detection constraints from Fermi-LAT \cite{Fermi-LAT:2015att}. The regions below the red line are allowed whereas the shaded region above it is disallowed from recent indirect detection constraints from Fermi-LAT.

In Fig. \ref{fig:5}, the permitted region lie below the red curve. It's noticeable that the DM mass range ($ M_{DM}\geq 300$ GeV) could actually yield the accurate relic abundance within our model and is also compliant with the bounds from DM annihilation. In regions of lower mass, the primary mechanism for DM annihilation is the s-channel process through Higgs mediation into SM fermions. We have identified that the key processes contributing to the relic abundance of DM include: $\chi_{1} \chi_{1}$ $\rightarrow$ $\tau^{+}\tau^{-}, W^{+} W^{-},h h, Z Z, b \bar{b}$. These annihilation channels play a significant role in generating the correct relic abundance for DM within this intermediate-mass range of the DM candidate. Consequently, the DM candidate $\chi_{1}$ in our model effectively satisfies both the constraints for relic abundance and direct detection cross-section within the mass range of $ M_{DM}\geq 300$ GeV.

Hence, from the above analysis, it becomes evident that a potential DM candidate within the mass range of $M_{DM}\geq 300$ GeV, introduces new particles that go beyond the SM. The incorporation of these new particles results in an interesting observation - their introduction leads to a positive vacuum behavior, extending even to the PLANCK scale $M_{Pl}$. These additional particles possess the capacity to induce modifications in the behavior of the quartic coupling $\lambda_h$. More specifically, they may contribute in a manner that guides $\lambda_h$ along a trajectory that maintains its positivity across the entire energy spectrum, extending up to the PLANCK scale $M_{Pl}$. This outcome represents a notable departure from the previous scenario and emphasises the captivating role these newly introduced particles play in upholding the theory's stability even at exceptionally high energy levels. This novel insight stands as a distinctive and significant contribution to this study.
\section{Conclusions}
\label{sec:5}

Of the many shortcomings of the SM, we revisited SM electroweak vacuum instability along with addressing the nature of intermediate mass range DM, in SO(10) GUT in  this work. At an energy scale of around $10^{10}$ GeV, the SM suffers from a vacuum metastability \cite{Degrassi:2014hoa, Rojas:2015yzm}. The addition of extra scalars could provide a way to stablize the vacuum up to the Planck scale. In this regard, we  delved into an exploration of a viable range of dark matter (DM) masses within a non-supersymmetric SO(10) Grand Unified Theory (GUT) scalar dark matter model. This framework encompasses a scalar singlet denoted as $S$, in conjunction with an inert doublet represented by $\phi$. Notably, these components are odd under discrete $Z_2$ matter parity $(-1)^{3(B-L)}$. The dark matter arises from the combination of the $Z_2$-odd scalar singlet $S$ and the neutral component of the doublet $\phi$, both belonging to a new scalar \textbf{16} representation inherent to SO(10).

We obtained regions in the parameter space which adheres to the observed DM relic density. Our investigation extends to the assessment of one-loop vacuum stability, achieved through the solution of Renormalization Group Equations (RGEs) governing the model's parameters. Subsequently, we rigorously evaluated the model's predictions, consistent with the boundaries of contemporary theoretical and empirical constraints. We successfully find a parameter range that concurs with considerations such as relic density, direct search bounds derived from the latest XENON1T findings, and the persistent stability of the electroweak vacuum, extending its validity up to the Planck scale. The novelty of the study is that it highlights a potential dark matter candidate which is a mixture of scalar singlet $S$ and inert doublet $\phi$ within a new intermediate mass range of 300$\leq M_{DM}\leq$1000 GeV. This region is shown to be consistent with the bounds from current direct detection XENON1T experiment \cite{XENON:2018voc} as well as indirect detection bounds of Fermi-LAT \cite{Fermi-LAT:2015att}. Additionally we have shown that unlike the SM case, the introduction of new particles intriguingly yields a positive vacuum behavior, extending all the way to the PLANCK scale $M_{Pl}$. These additional particles possess the capacity to induce alterations in the behavior of the quartic coupling $\lambda_h$. Specifically, they can contribute in a way that steers $\lambda_h$ along a trajectory ensuring its positivity across the entire energy spectrum, reaching up to the PLANCK scale $M_{Pl}$.

Thus, the prediction of mixed scalar DM in the intermediate mass range and upholding the vacuum stability at high energy from within the same SO(10) GUT group constitutes a distinctive and substantial contribution of this study. The predictions stemming from this model hold promise for potential validation in upcoming highly sensitive experimental searches focused on DM such as XENONnT \cite{XENON:2024wpa}. Moreover, these predictions align smoothly with the beauty found in grand unified theories.

\section{Acknowledgment} 
We acknowledge the RUSA and FIST grants of Govt. of India for support in upgrading computer laboratory of the Physics Department of Gauhati University, where this work was completed.

\end{document}